# Designing electrostatic MEMS-based electron optics: the case of the spiral phase plate

P. Habibdazeh Kavkani[1*], A. Tavabi[2*], P. Rosi[3], A. Roncaglia[4], E. Rotunno[3], L. Belsito[4], S. Sapienza[4], S. Frabboni[1,3], R. Dunin.Borkowski[2], M. Beleggia[1**], V. Grillo[3]

1. Department of Physics, Informatics and Mathematics, University of Modena and Reggio Emilia, 41125 Modena, Italy
2. Ernst Ruska-Centre for Microscopy and Spectroscopy with Electrons and Peter Grünberg Institute, Forschungszentrum Jülich, 52425 Jülich, Germany
3. CNR-Nanoscience Institute, S3 center, 41125 Modena, Italy
4. CNR-Institute for Microelectronics and Microsystems, 40129 Bologna, Italy

*Equal contribution

**Corresponding author

**Abstract**

A new generation of microfabricated MEMS for electron optics is changing electron microscopy for the better. These devices allow operations on the electron beam that are impossible with conventional electron optics. Unprecedented phase landscapes like tunable spiral phase plates and localized strong phase gradients are just some examples of what can be achieved. This work establishes the methodological foundation to design and control MEMS based phase plates. The design strategy is rooted on a novel analytical and numerical modeling of thin electrodes with accurate account of the fringing fields having a major role in the thin-MEMS geometry. We designed, fabricated and characterized experimentally a spiral phase plate, and assessed the quality of the generated vortex beam while discussing the most relevant control parameters and design approaches.

**Introduction**

Electron microscopy is undoubtedly one of the most versatile techniques to characterize materials and biomaterials. Underpinning its success is the ability to build lenses and more complicated electron optical systems than what is possible in light optics.

Today, a variety of electron-optical systems are available to correct high-order geometrical aberrations [1–4] and chromatic aberration [5,6], to provide magnetic field-free atomic resolution [7] and to push energy resolution into the meV range [8]. These opportunities, however, come with a cost: the increased size and complexity of the electron microscope.

Yet, the upcoming revolution in electron beam control may not arise from large optical systems but from the introduction of miniaturized electron optics based on MEMS technology [9,10]. In particular, thin Si-based chips can be placed in various aperture planes of the microscope and used to affect the phase of the electron beam.

While nothing can arguably improve upon the stability and precision of a large magnetic lens, miniaturized electron optics use magnetic and electric fields generated in a region very close to the beam to produce more complex phase patterns.

Thanks to the small gap between the electrodes and the beam, miniaturized optics can produce large phase effects with moderate voltages or currents. However, the most important advantage of miniaturized optics is the opportunity to modulate the electron phase via external control of the MEMS device.

Recently, the realization of a new electron optical configuration called "Orbital Angular Momentum (OAM) sorter" has represented a tangible demonstration of the fruitful complexity that can be achieved by MEMS based optics [9].

The device combines two MEMS phase plates to modulate the electron phase over two subsequent optical planes. Their collective effect can be described as a conformal mapping of the electron wavefunction. The OAM sorter has allowed for the first time to measure in a single image the orbital angular momentum spectral components of an electron wavefunction. Later, it provided us with the very first OAM-resolved EELS spectrum [11]. More complex conformal mapping schemes are envisioned to improve the quantitative performance of electron microscopy [12,13].

The counterpart of this complexity is the requirement of sophisticated controllers such as artificial intelligence systems that can automatically align and optimize complex and imperfect optical systems with many control parameters (for example the set of electrodes voltages). These systems need to be integrated into microscope operations [14].

The envisaged flexibility of MEMS phase plates, however, has a fundamental limitation: the field equation of Electrostatics combined with the many geometrical constraints imposed by the instrument restrict our ability to modulate the electron phase on a point-by-point basis.

As described in Ref. [9], our MEMS electron optics is based on very thin lenses that, under realistic circumstances, such as well-behaved boundary conditions and charge neutrality, can only introduce harmonic phase modulations to the electron wavefunction. In this context, harmonic means that the phase distribution imparted onto the beam by the device satisfies the 2D Laplace's equation: $\nabla_{xy}^2 \varphi = 0$ where *x,y* are the coordinates of any plane orthogonal to the beam propagation direction *z*, and $\varphi$ is the phase of the electron wave function. It is Laplace's equation, not Poisson's, because the beam should not propagate through solid portions of the device, and, therefore, no charge intersects with the beam path. Such unwanted interaction, in fact, would bring along charging and fast aging of the device, as well as detrimental amplitude modulations of the electron wavefunction (loss of coherent electrons). Only minor deviations from this rule may be tolerated if the few electrodes along the beam path are small compared to the beam size. We shall name phase plates of this sort "quasi-harmonic".

The alternative in the field of MEMS technology is the so called pixelated approach, where individually addressable electrodes (acting similarly to the pixels of a spatial light modulator) modify the electron phase in a small and confined region, thereby producing a discrete approximation of an arbitrary phase landscape. All this is achieved on a very limited number of pixels (up to 48 with current technology) and with a large filling factor [15–17]. The use of these phase-plates is potentially different from the harmonic phase plates we discuss here For example they are potentially well suited for simplified single element spherical aberration correctors and highly irregular phase landscape.

Harmonic phase modulations are relevant as they enable conformal transformation of the wavefunction [12]. Also, because single harmonic or quasi-harmonic phase plates can be used to generate interesting wave forms [18,19]. One of these is the spiral phase plate that generates electron vortex beams and that can also be used post-specimen on diffracted electrons for edge contrast enhancement [20].

The aim of this paper is to devise a general design strategy for MEMS-based electron optics considering current technology limitation and geometric effects by looking at the specific use-case of spiral phase plates. Various designs for spiral phase plates have been realized in the past with magnetic and electric fields [21,22]. In the case of electrostatic designs both analytical and numerical approaches are used here to establish a relation between the applied voltages in the thin MEMS plane and the phase change experienced by the electron beam.

The calculations based on both the above methods allow us to account for and compensate for the thin electrode geometry and the potentially large fringing fields it creates. Moreover, these approaches can be used in great generality to generate any form of harmonic phase landscape. Experimentally, we can reproduce the near-exact phase landscape of the spiral phase plate and control some of its main aberrations by controlling the boundary conditions. We also demonstrate that the potentially complex boundary conditions can be established by a limited number of electrodes using a controlled current flow (horizontal and therefore not affecting the phase) and a set of resistive paths that introduce a potential gradient on the boundary electrodes.

To treat the phase introduced by MEMS based phase plates we need to solve in general the 3D electrostatic problem given a series of electrodes whose bias can be controlled externally.

In fact, the phase impressed to the electron is related to the potential by the equation $\varphi = C_E \int V dz$ where $C_E$ is the appropriate electron coupling constant [23,24], and the integral is taken along an electron trajectory parallel to the optic axis of the microscope and extending above, below, and within the phase plate active region.

In our geometry the MEMS comprises of a conductive layer (typically doped Si) with a thickness $t$=10-30 μm, with a thin layer of gold thermally evaporated over it. As we want the electron beam to interact as little as possible with matter along its path, we create a circular hole with little (only two thin electrodes in this case) or ideally no material in the way of the electrons.

The details of a typical structure are summarized in Fig 1. Fig 1a) shows a low magnification 3D sketch of the device where the total thickness is typically in the range of 2-300 μm. Fig 1b) displays a zoomed-in sketch of the device active area, where beam electrons fly by and interact; highlighted are the main materials and the layered structure of our MEMS phase plate. Fig 1c) portrays a 3D representation of how the phase plate re-shapes the electron beam wavefunction

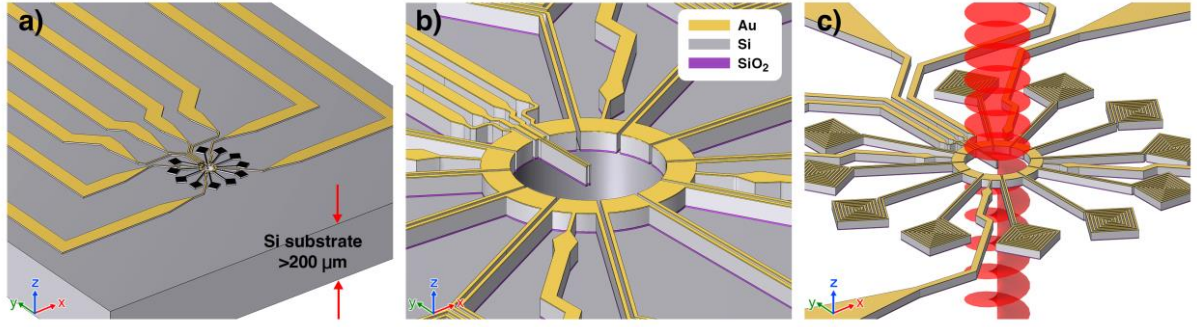

*Fig 1 Typical structure for MEMS technology with connection from pad to active area ( a). The active area (b) has a vertical hole for electron beam to pass through. In (b) it is also possible to observe the Si on insulator technology that allows for trench isolation. Fig (c): schematic representation of how the spiral phase plate works, where an incoming plane-wave wavefront is reshaped into a spiraling wavefront due to the phase shift induced by the device*

In all physical scenarios where we assume overall charge neutrality the phase modulation of the electron wave function $\varphi$ that originates from a set of fixed charges satisfies the equation [25]:

$$\nabla^2_{xy}\varphi = \sigma \qquad [1]$$

where $\sigma$ is the projected charge and $\nabla^2_{xy}$ is the 2D Laplacian operator acting over a plane (x,y) perpendicular to the optic axis z. Since $\sigma \neq 0$ only wherever materials intersect the electron beam path, this leads directly to the above-mentioned harmonic constraint of the phase plate. Quasi-harmonic phase plates are those that contain small regions where $\sigma \neq 0$.

In the vacuum surrounding the electrodes the 3D Dirichlet problem for the potential *V* is well defined:

$$\begin{cases} \nabla^2 V = 0 \\ V|_{\partial\Omega} = v(s) \end{cases} \qquad [2]$$

where $v(s)$ is the potential distribution on the surface $\partial\Omega$ of the electrodes occupying a volume of space $\Omega$ parametrized by a set of curvilinear coordinates $s\epsilon\partial\Omega$. We assume that the potential at infinity is well behaved. By integrating along the optical axis, we can write the analogue 2D Dirichlet problem for the phase in a circular aperture $\Sigma$ of radius *R* surrounded by electrodes as

$$\begin{cases} \nabla^2_{xy}\varphi = 0 \\ \varphi|_{\partial\Sigma} = f(\theta, \rho = R) \end{cases} \qquad [3]$$

In eq. 3 we used the polar coordinates system $\rho, \theta$, and $\partial\Sigma$ represents the edge of the aperture, along which $f(\theta, \rho = R)$, an arbitrary function, is the assigned boundary condition.

In the general case we do not know the phase value at a given boundary position, here parametrized by the azimuthal variable $\theta$. This is because it depends on the full 3D development of the electrostatic potential, including especially its fringing fields protruding in the vacuum above and below the electrodes that have a total thickness *t*.

We focus our study on the case of a spiral phase plate, which is harmonic almost everywhere except for a small line of discontinuity produced by two lines of charge.

In order to develop an analytical solution, we will treat here the completely harmonic case and add the non-harmonic corrections separately. If the two phase contributions from the harmonic and non-harmonic potentials are matched at the boundary, the unicity of the solution assures us it is the correct solution.

In the case of the vortex beam, the analytical solution for the non-harmonic part is a set of two parallel lines or vertically aligned thin parallel slabs of opposite charge that reach for the center of the beam [22]. This translates to a pair of smooth slightly tapered electrodes in the actual implementation, where the shape is derived from equipotential surfaces in 2 dimensions [25]. These electrodes are here referred to as "chopsticks".

The harmonic contribution to the phase can be described by a multipolar decomposition

$$\varphi(\rho, \theta) = \sum_n a_n \rho^n \sin(n\theta + p_n) \qquad [4]$$

where $a_n, p_n$ parametrize the multipoles amplitude and orientation.

We define in particular $a_n = A_n$ the coefficients for the ideal phase that we aim to reproduce (in this case $\varphi_{aim} = \ell\theta$ where $\ell$ is the number of angular momentum quanta of the vortex beam). For an ideal spiral phase plate oriented in such a way that $p_n = 0 \; \forall n > 0$, we have:

$$A_n = \frac{\ell}{2\pi} R^n \int_0^{2\pi} \theta \sin(n\theta) \, d\theta = -\frac{\ell}{n} R^n \qquad [5]$$

In an ideal phase plate the $A_n$'s would be directly the multipole coefficients if we could neglect the fringing field (SCOFF: sharp cut off approximation [26]).

The analytical calculation described in Appendix 1 exposes the connection between phase and potential multipole coefficients as a function of the electrode thickness. The appendix indeed uses the relation for each multiple between the cases with and without the fringing field effect. Such correction is increasingly small for increasing order of multipole since qualitatively what matters is the distance between neighboring poles with opposite polarization. Such distance geometrically decreases as 1/n.

This leads to determination of what potential profile the boundary electrodes must have

$$V(R, \theta) = \frac{1}{C_E t} \sum_n \frac{A_n}{1 + \frac{R}{cnt}} \sin(n\theta) \qquad [6]$$

which we can establish in practice by controlling externally each of the available electrodes. In eq. (6), $c$ is a coefficient of the order of unity that is approximated by $c=\pi/2$ in the simplified calculation in Appendix 1; its true value should be determined once for each device by a set of more realistic simulations. The interpretation of equation (6) is the following: since the desired harmonic phase landscape is described by the sum of many multipoles of the form of coefficients $A_n$, the multipole decomposition should hold for the actual 3D potential in cylindrical coordinates. However, each multipolar component develops different extensions of the fringing field that must be accounted for when transitioning from the 3D voltage to the phase problem.

In the Appendix we develop a full solution of the fringing field generated by a cylindrical-shell electrode biassed controllably around its inner surface. Such solution is derived starting from a fixed charge distribution on the electrode, which yields a potential that varies along *z* (hence, it is not equipotential). The charge distribution is then relaxed, and the resulting equipotential is approximated by the averaged potential from the fixed charge distribution. As shown explicitly in the Appendix, this approximation yields the correct structure (thickness and multipole order dependences) of the potential multipole coefficients as in eq. (6), but appears to overestimate the coefficient *c*.

The correction vanishes if *n* is large or if *t>>R* i.e. when the electrodes are so thick that the fringing field can be neglected.

The same solution (see eq. (6)) for the set of biases for a specific device geometry can be found also numerically by using COMSOL finite element software. With it, the user can require the minimization of a functional potential *U* with respect to some control parameters that, in this case, are the applied biases on some lateral electrodes.

The function to be minimized can be written as a linear combination of the phase from each electrode biased separately. As minimization domain we choose a circumference $Crc$ with radius ≈0.9*R* so that it is mainly affected by the boundary conditions

$$U(V_{i,bias}) = \left| \int_{x,y \in Crc} [\varphi(V_{i,bias}) - \varphi_{aim}] dx dy \right| \qquad [7]$$

For example, if considering only the multipolar boundary condition and the two conducting lines

$$U = \left| \int [\sum_n b_n \rho^n \sin(n\theta + p_n) + \sum_m \varphi_m - \varphi_{aim}] dx dy \right| \qquad [8]$$

where *n* as above indexes the multipoles and *m*=1,2 are the bias on the two chopsticks. Since the problem is linear the solution quickly converges to a stable charge neutral solution.

Using COMSOL we can perform the minimization with more general electrodes geometries. In supplementary A5 we also calculated the exact solutions considering very realistic MEMS geometry and grounding, while here for a better comparison with the analytical case consider only isolated conductors in a ring.

Figure 2 shows the results of the analytical and numerical calculations for the boundary condition $V(\theta, \rho = R)$, examining different values of the electrodes thickness in the beam direction.

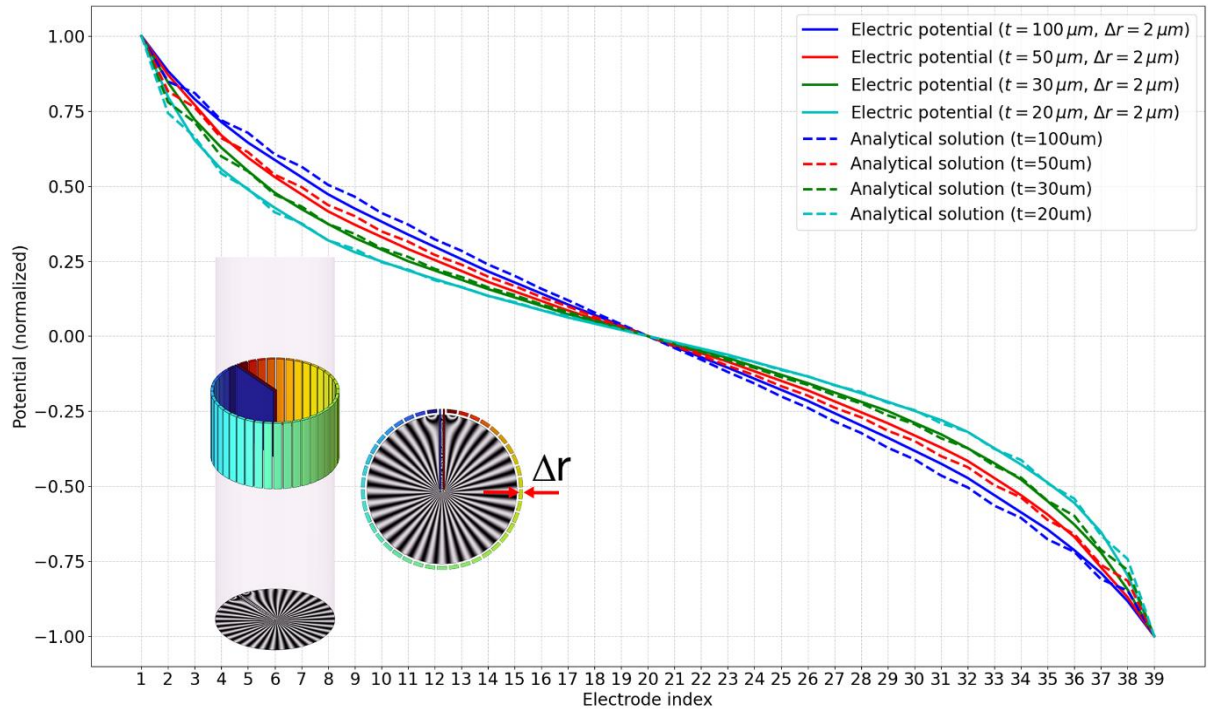

*Fig 2 Model for the optimization of boundary conditions electrodes with analytical and numerical (finite elements) results using a large number of adjustable electrodes. For consistency with the analytical model, we considered a very thin ring of radius R=40 μm. In order to obtain a vortex boundary condition we need an azimuthally linear phase ramp. This is obtained with the complex profile of potential applied to the electrodes as described in figure (the potentials are normalized to the maximum). The inset shows the resulting phase contour map $\sin(N\varphi(x,y))$ with N an appropriate amplification factor.*

Whereas the target phase profile near the boundary should just be an azimuthal linear ramp, the applied potential turns out to be non-linearly dependent on the azimuth θ, with a very small voltage variation in front of the chopsticks and a very steep gradient in their proximity.

The similarity between the analytical and the numerical solutions validates our method to predict the compensation of the biases. Minor differences arise from the approximation in the theoretical model (i.e. neglecting the induction between the chopstick and the boundary electrodes), and in the detailed description of the chopstick in the numerical model (e.g. size and sampling of the mesh in the COMSOL model).

For example, here for the very thick electrodes case we can see small deviations from the analytical model that could be solved with a larger COMSOL model with cell boundary at larger distances.

The resulting phase as *z* integral of the potential is everywhere close to the ideal linear azimuthal ramp even if in the plane of the device the functional dependence of the applied potentials on θ is significantly different.

The analytical methods permit us to interpret the correction to the applied potential in terms of multipoles and therefore also of elements of microscope optics while the finite elements solution is more general and detailed. Finite elements simulation permits, for example, to introduce more realistic elements like the grounded substrate or even the MEMS holder (see supplementary A5) but the same solution can be achieved in the analytical method by using an effective value of the constant *c* in Eq. 6. Finally, a realistic COMSOL model can also account for finite resistance of the circuit leading to the electrodes, but we did not do this systematically.

Our present technological solution is limited to 8 contacts, which clearly limits the solution due to the low number of independently biassable electrodes. Even with future designs and implementations there will always be a limitation on the number of available connections. For this reason, we adopted a different solution: a new geometry that employs resistive paths between connections and current flow to increase the number of active electrodes at the device boundary. In this way intermediate electrodes that are not directly controlled by the external bias reach an intermediate potential between the directly controlled biases. With a clever use of this trick, we are able to address a total of 14 electrodes including the chopsticks protruding in the center of the beam.

The multipolar nature of the correction and the use of intermediate electrodes to account for the limited number of contacts can be taken into account when designing the real device. Considering both the multipole expansion and the nature of the applied potential in real space the device can be designed in two different ways:

1) **Real space approach:** we match directly the azimuthal shape in real space by appropriately positioning a number of independently controlled electrodes along the boundary. We concentrate on controlling elements in the angular region with the steepest gradients.
2) **Multipole space approach:** we start from an ideal azimuthal ramp and decompose the deviation between the ideal and real boundary conditions by adding multipoles of increasing order. The control electrodes are positioned according to the symmetry of the multipoles to be controlled.

The design approach 1 (fig 3a) is made possible by the fact that the solution is linear in a large portion of the angular range. A linearly decaying potential is now possible with the use of current and resistive paths and will be described better in the experimental section.

The design approach 2 (fig 3.b) is explicitly relying on the multipole decomposition that is of general applicability. In this case, the electrodes are disposed periodically along the boundary with periodicity dictated by the multipole that needs to be corrected (in this case quadrupole and hexapole). Since only low order effects can be controlled with a finite number of controls, we cannot control strong gradients.

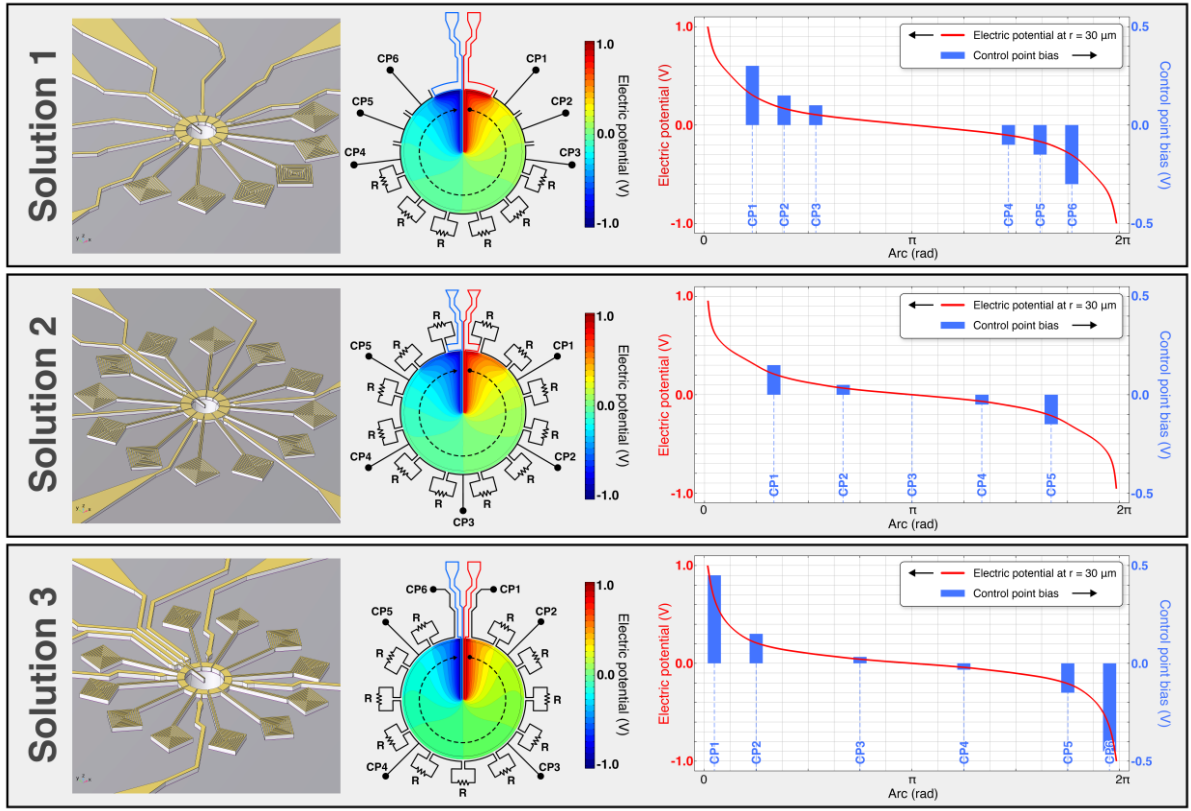

*Fig 3 Different solutions for the spiral phase plate with different ideas for the boundary condition electrodes. The position of the directly controlled electrodes is distributed according to different strategies: Solution1 (up) is based on real space approach (i.e. taking advantage of largely linear area in the required bas in front of the chopstick); Solution 2 (center) highlighting the multipolar character (i.e., controlling the hexapolar term); Solution 3 (Bottom) mixing together both approaches (i.e., highlighting the importance of the discontinuity close to the chopstick but also controlling at least the quadrupole term). The image reports a schematic of the role of the resistors (labyrinth-like structure) and of the independent control points. These are also reported in the azimuthal graph in the last column that shows the ideal boundary profile.*

Fig. 3 summarizes examples of the different approaches: solution 1in fig 3 concentrates all the control electrodes in the region close to the chopstick assuming a linear gradient in the whole frontal region. This is what could be called a real space approach because it is based on the knowledge of the specific solution with small gradient. Solution 2is mainly based on hexapolar geometry that allows to directly control the 3-fold astigmatism (i.e., the 3$^{rd}$ order multipole). This geometry is an explicit example of the multipole space approach, and it allows a perfect correction of the first two multipoles with the higher order multipole corrections supposed negligible, as the $\frac{1}{1+\frac{R}{cnt}}$ term in eq. (6) quickly reduces to 1 for large *n*.

Solution 3 is a compromise between the previous solutions. It uses the real space knowledge that there is a very steep gradient in proximity of the chopsticks so that 4 control electrodes are concentrated only in that region while the remaining are nearly evenly spaced in the remaining angular regions. This permits us to control the two dipole and one quadrupole term while renouncing the control of the hexapole term. We focus here on the type (c) approach as it gives the best compromise between the two above design strategies and should perform slightly better according to simulations.

Experimentally we worked on two FEI-TITAN ("TITAN T") microscopes operated at 300 keV. The system is not endowed with imaging corrector and an aperture with radius 20 um is used to reduce the effect of the spherical aberration of the imaging systems. We also performed large area off-axis electron holography on a Thermofisher Talos operated at 200 keV and equipped with a biprism to measure the phase around the electrodes. On the same instrument, we recorded the small angle diffraction pattern to observe the shape of the vortex beam.

The MEMS were mounted on a special Thermofisher Scientific aperture-holder with 8 pass-through contacts. The contacts were externally controlled trough house-made electronics (that we called "MINEON") that can be directly driven by a PC using either a custom python code or a Labview interface.

Fig 4a shows the geometry of the electrodes where the resistive paths have a "labyrinth-like" geometry. Measurements indicate that each labyrinth corresponds to a resistor of 50-100 Ω. As the current is typically limited to a few mA we cannot apply more than a few volts of potential difference between neighboring electrodes. However, this is acceptable for the purpose of generating vortex beams of a few tens of OAM quanta, because the only large potential gradient is concentrated over the two non-connected electrodes forming the chopstick.

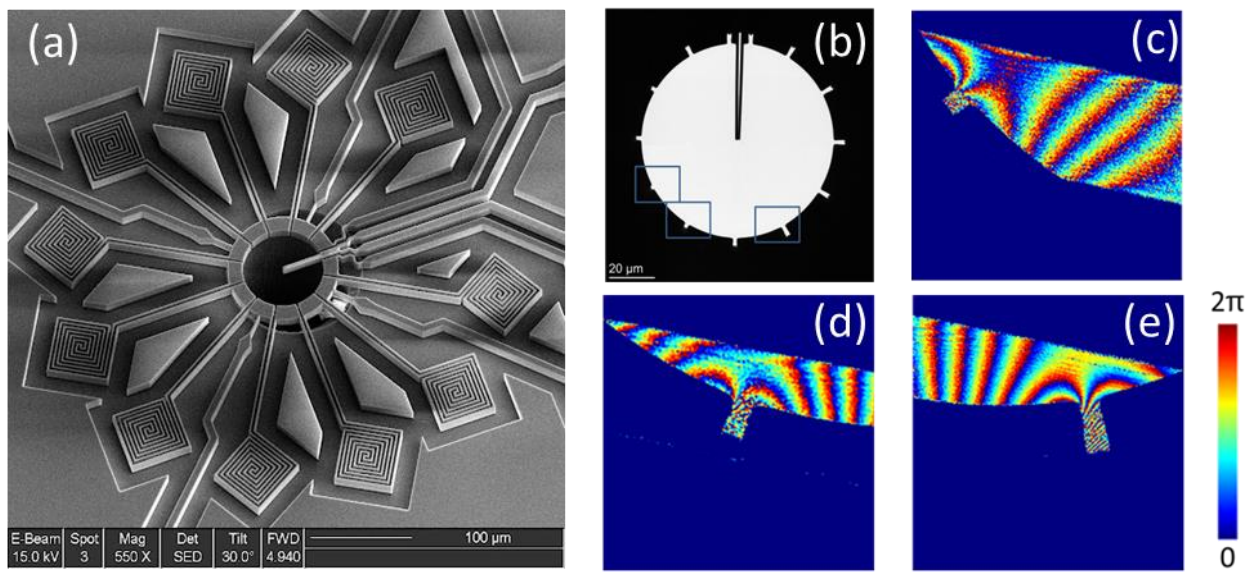

*Fig 4 (a,b) SEM and TEM image of a MEMS device (c,d,e) holographic phase reconstruction in regions nearby adjacent boundary electrodes. The regions from which these were taken from are highlighted in (b). The boundary electrodes are connected through a resistive path where the flowing current determines a discrete potential drop. In addition to the clear phase discontinuity a linear phase gradient is present because of unavoidable perturbation of the reference wave.*

The phase images shown in fig 4c-e were acquired with off-axis electron holography and indicate the net change of potential at the separation between two consecutive boundary electrodes when a current flows between them. The position of the analyzed areas is indicated in the schematic image in Fig. 4b with blue boxes. The data confirms that we can control the potential on the boundary electrodes to have a stepwise potential drop. In fact, we measure a dipole–like phase gradient between two electrodes.

Next, we tested the realization of a large vortex (angular momentum $\ell \approx 50$) in low angle diffraction mode and verified the effect on the beam shape of the different boundary conditions obtained by modifying the bias distribution at the contact points.

The potentials are plotted starting from the first chopstick electrode and ending to the opposite one along the azimuthal direction. A visual inspection of the various images and potential profiles indicates that a near-ideal vortex phase can be achieved with a curve that resembles a discrete version of the theoretical one shown earlier in Fig. 2 with the addition of some offset bias on both sides of the chopstick.

To be more quantitative, we introduced a quality factor C that defines the "roundness" for the intensity profile. The definition of C is inspired by the zero-th order term digital OAM decomposition for a completely flat-phase ring beam [27] and reads as follows

$$C = \int |\int I(\rho,\theta)\rho \exp(ik\rho)\, d\theta d\rho|^2 dk \qquad [9]$$

where $I(\rho,\theta)$ is the intensity expressed in polar coordinates (centred at the intensity barycentre) and k is an integration variable related to the radial variable. With appropriate normalisation C should be 1 for perfectly circular ring and strongly reduce to less than 0.2 for strongly deformed probe. The values of C have been reported on each beam shape.

The electronics allowed us to control each of the 8 contacts separately or to group them with a few shape description parameters. The parameters are described in supplementary.

Some shape parameters like curvature and scale factor between the chopstick, and the boundary electrodes are more relevant than others.

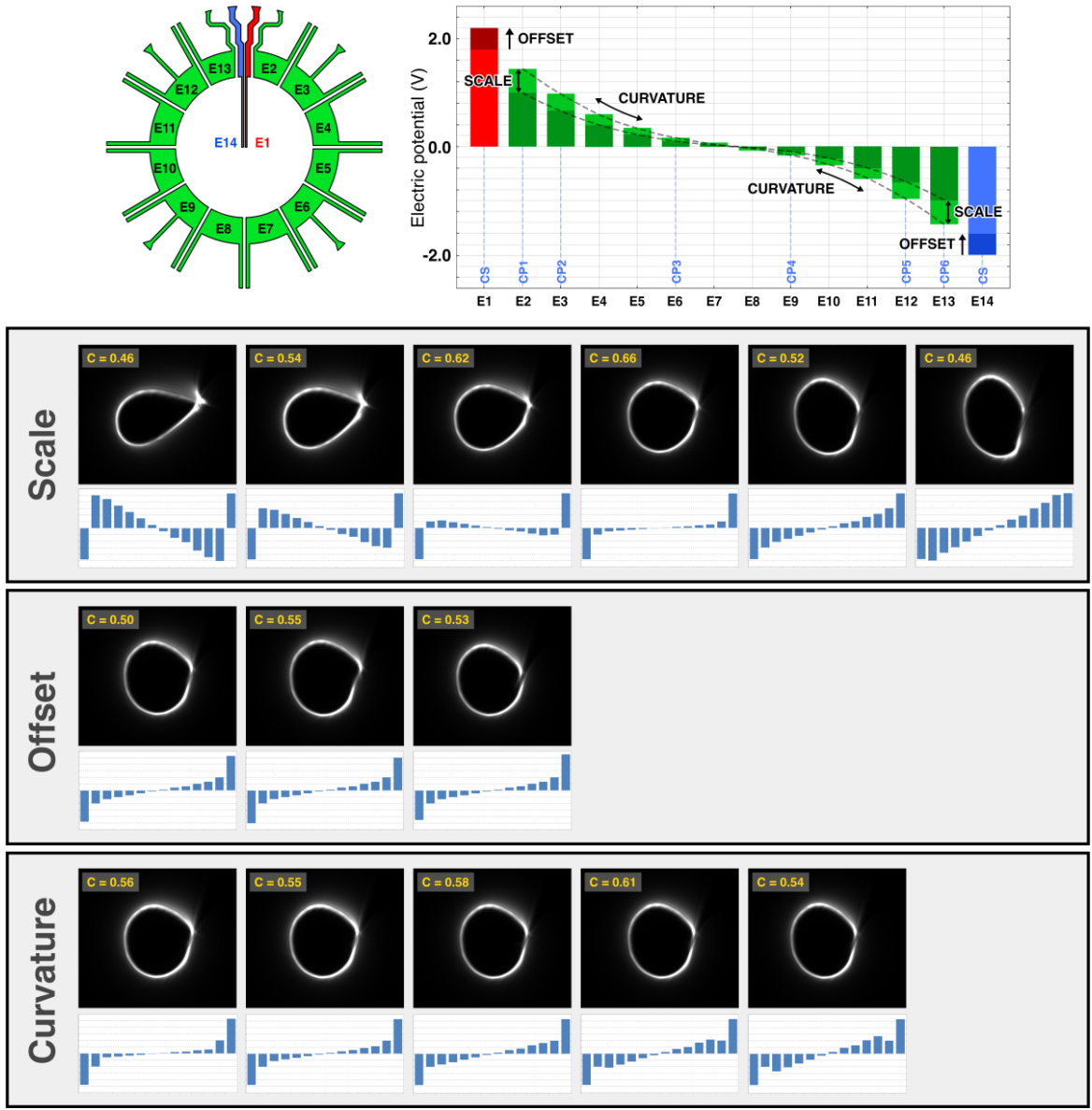

*Fig 5 Optimization test for the beam shape. We defined 3 different shape factors as described in top scheme (and in appendix) and vary systematically their value and observe the different deformations of the vortex. The applied electrode potentials distributions in absolute units are shown below each image. In the first row we change the scale factor of the boundary electrodes with respect to the main chopstick electrodes. In the second row we add an offset potential equal for both chopstick elements: it has the effect for example to compensate charge effect that accumulates at a tip because of the electron beam and for the not ideal shaping of the electrode. This parameter can be adjusted to control discontinuity. In the last row we fine tune the quadratic terms controlling the deviation from linearity introduced by taking into account the finite thickness of the device.*

As a further test of the analytical model above we were able to correct most of the boundary condition using the aberration control of the microscope. Fig 5 shows how a completely wrong potential profile on the boundary electrodes produces a deformation that elongates the vortex. After the 2-fold astigmatism correction the beam shows clearly the residual 3-fold astigmatism that can be compensated with the hexapoles of the image corrector. This indicates also that the boundary electrodes (without the "chopstick") could be used to control both 2-fold and 3-fold astigmatism in common microscope operations or even be combined in a smaller version of spherical aberration correction optics.

It is worth stressing that the boundary electrodes alone would never be able to create a vortex phase without the two chopsticks. The vortex phase implies a discontinuity in the potential.

In fact, a full aberration corrector has been used to approximate the vortex phase only in a limited radial range and the result was relatively poor [28]. On the other hand, the presence of the chopstick always limits the continuity of the wavefunction and the purity of the OAM state: all diffraction images feature a dimmer point corresponding to the chopstick shadow.

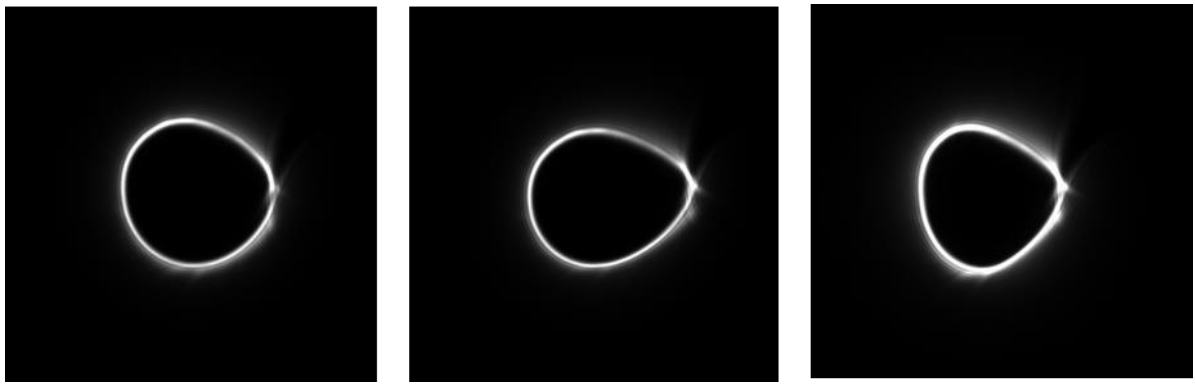

Fig 6 Experimental demonstration that varying the boundary electrodes bias mainly corresponds to a combination of 2-fold and 3-fold astigmatism. When from a well tuned beam (left) the scale factor of the first raw of fig 5 is varied a squished circle is obtained (center figure) and once 2-fold astigmatism is corrected with the microscope quadrupoles a clear effect of residual 3-fold astigmatism is visible (right).

In conclusion we demonstrated that we can produce a complex phase landscape with large accuracy by using analytical and numerical methods for the modelling and design, nanofabrication for the realization of complex electrodes structures that feature ohmic resistors in order to drive a larger number of electrodes with a finite set of contacts in the holder. The multipole decomposition of the boundary conditions is particularly useful to control the strong fringing field effects on them allowing near-arbitrary control on quasi-harmonic phase plates.

This smart organization of electrodes and a good design can be used to overcome the shortcomings of the limited number of control points, of the presence of fringing fields and even of microscope aberrations. This is therefore a very important step in the realization of MEMS based custom phase elements for electron optics in electron microscopes and charged particles optics alike.

**Acknowledgment**

We acknowledge support from Ministero dell'Università e della Ricerca (PRIN Project No. 2022249HSF "AI-TEM") and we acknowledge support from the European Union's HORIZON EUROPE framework program for research and innovation (Grant Agreement No. 101094299 "IMPRESS").

## Supplementary

### A1 Multipole expansion

The consequence of the hole circular geometry is that the solutions of the phase problem (typically considered for r<R) "thick electrode" for which fringing field are negligible (e.g. t>>R) and "thin electrode" can be all written as:

$$\varphi_{thin}(\rho, \theta) = \sum_n a_n \rho^n \sin(n\theta + p_n) \tag{A1}$$

This is a consequence of the harmonic constraint that in both cases induces a relation between the radial and azimuthal part. The "thick electrode" case will be written hereafter as:

$$\varphi_{thick}(\rho, \theta) = \sum_n A_n \rho^n \sin(n\theta + p_n) \tag{A2}$$

(see eq. (4) in the main text)

The thick electrode case is equivalent to assuming there is no fringing field. Therefore, the ideal phase landscape is produced with no corrections needed between potential and phase multipole coefficients. In fact, since the coefficients necessary to reproduce a target phase distribution are:

$$\varphi_{thick}(\rho = R, \theta) = \sum_n A_n R^n \sin(n\theta + p_n) \tag{A3}$$

and with thick electrodes the potential does not depend on the axial coordinate $z$, the boundary phase $\varphi_{thick}(R, \theta) = C_E V(R, \theta) t$ is directly related to the applied bias to the boundary electrodes.

For the thin electrode case the problem of finding the relation between phase and applied bias is transformed in the problem of finding the relation between $A_n$ and $a_n$. For this, we need to analyze the actual electrostatic solution.

The effect of the fringing field and therefore the relation between the two cases has been calculated in the following appendix A2. We find in agreement with many partial results in the literature that each multipole contribution of the thin electrode case is mapped onto the same-order multipole contribution of the thick electrode case evaluated at the effective thickness $t_n = t + \frac{R}{nc}$ where $c$, a factor on the order of unity, is necessary to account for the exact extension of the fringing field depending on geometry of the electrodes.

Appendix A2 indeed clarifies that higher order multipoles have a much shorter fringing field. The relation between the coefficients is summarized as:

$$a_n \approx A_n \frac{t + \frac{R}{n}}{t} = A_n \left(1 + \frac{R}{cnt}\right) \tag{A4}$$

The thin sample tends to enhance the phase with respect to the thick case. Therefore, we must correct the potential coefficients when relating the thick case to a thin case according to:

$$A_n^{corr} = A_n / \left(1 + \frac{R}{cnt}\right) \tag{A5}$$

This formula is a universal rule on how to create boundary conditions for any case of harmonic phase profile projected on a circular hole. In fact, the phase profile at the frontier of the hole can be predicted from the desired analytical phase expression and the coefficient $A_n$ arises from the Fourier decomposition in the angular basis.

We can therefore pretend to be in the thick case and apply a voltage bias on electrodes:

$$V(R,\theta) = \frac{1}{C_E t}\sum_n \frac{A_n}{1+\frac{R}{cnt}} R^n \sin(n\theta + p_n) \quad [A6]$$

The correction vanishes if *n* is large or if *t>>R*, i.e. the thick electrode case. The interpretation is the following: if the desired harmonic phase landscape is described by coefficients $A_n$ the applied bias should have the form of eq. 6.

The terms $p_n$ reflect only the orientation of the desired phase landscape and are not altered by the fringing field. For convenience it will be dropped hereafter.

We can also rewrite the desired potential at the boundary as the sum of the bias applied to n-order multipoles

$$V(R,\theta) = \sum_{n=1}^{\infty} V_n \sin(n\theta) \quad [A7]$$

Where now can be written in terms of $A_n$

$$V_n = \frac{1}{C_E t} A_n^{corr} = \frac{1}{C_E t} \frac{A_n}{1+\frac{R}{cnt}} \quad [A8]$$

**A2 Multipolar fringing field**

Let us consider a cylindrical shell *S* of radius *R* and height *t=2d* representing the active area of our phase plate, as sketched in fig. A2.1. The shell is externally biased to a known potential distribution as in equation A8 above. Such potential is established by an unknown charge distribution on the shell σ(*θ,z*) that we also expand in multipoles

$$\sigma(\theta, z) = \sum_{n=1}^{\infty} \sigma_n(z) \sin n\theta \quad [A9]$$

The potential in the shell interior *r<R* can be expressed as

$$V(\mathbf{r}) = \frac{1}{4\pi\varepsilon_0} \iint_S \frac{\sigma(\mathbf{r}_S)}{|\mathbf{r}-\mathbf{r}_S|} d^2\mathbf{r}_S \quad [A10]$$

where **r**=(*r,θ,z*) and **r**_S=(*R,θ_S,z_S*) are the position vectors of a generic point in the interior and on the shell, respectively.

By using the expansion of 1/*r* in cylindrical coordinates:

$$\frac{1}{|\mathbf{r}-\mathbf{r}'|} = \sum_{m=-\infty}^{\infty} e^{im(\theta-\theta')} \int_0^{\infty} J_m(kr) J_m(kr') e^{-k|z-z'|} dk \qquad [A11]$$

and considering that d²**r**<sub>S</sub>=Rdθ<sub>S</sub>dz<sub>S</sub>, we rewrite the potential as

$$V(\mathbf{r}) = \frac{R}{4\pi\varepsilon_0} \sum_{n=1}^{\infty} \sum_{m=-\infty}^{\infty} e^{im(\theta-\theta_S)} \int_0^{\infty} J_m(kr) J_m(kR) e^{-k|z-z_S|} dk \int_0^{2\pi} \sin(n\theta_S) d\theta_S \int_{-d}^{d} \sigma_n(z_S) dz_S$$

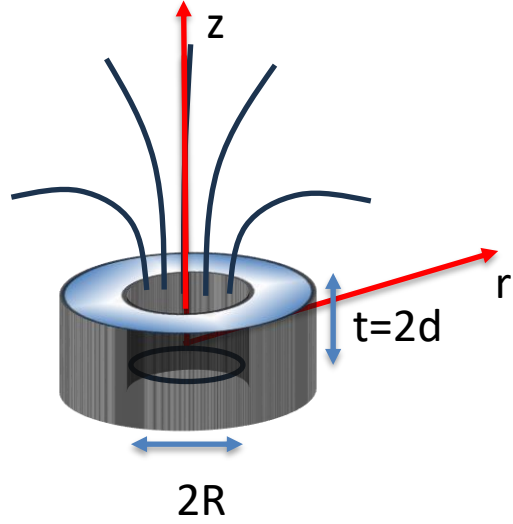

**Figure A2.1:** *Sketch of the geometry considered to represent the active area of the phase plate: a cylindrical shell of radius R (the shell is infinitely thin) of height t=2d.*

After carrying out the angular integration

$$\int_0^{2\pi} e^{-im\theta_S} \sin(n\theta_S) d\theta_S = i\pi(\delta_{m,-n} - \delta_{m,n}) \qquad [A12]$$

we obtain

$$V(\mathbf{r}) = \frac{R}{2\varepsilon_0} \sum_{n=1}^{\infty} \sin(n\theta) \int_0^{\infty} J_n(kr) J_n(kR) dk \int_{-d}^{d} e^{-k|z-z_S|} \sigma_n(z_S) dz_S \qquad [A13]$$

The application of the boundary condition

$$V(R,\theta,|z|<d) = \frac{R}{2\varepsilon_0} \sum_{n=1}^{\infty} \sin(n\theta) \int_0^{\infty} J_n^2(kR) dk \int_{-d}^{d} e^{-k|z-z_S|} \sigma_n(z_S) dz_S = V(\theta) = \sum_{n=1}^{\infty} V_n \sin n\theta \quad [A14]$$

allows us to link the known potential coefficients with the unknown charge densities

$$V_n = \frac{R}{2\varepsilon_0} \int_0^{\infty} J_n^2(kR) dk \int_{-d}^{d} e^{-k|z-z_S|} \sigma_n(z_S) dz_S \qquad [A15]$$

but leads no further since no family of orthogonal functions appears readily applicable to resolve the charge density profiles. Since, however, we are focusing our attention on the phase shift, we can make two steps back and project the potential by integrating the expression (A13) along the optic axis:

$$\varphi(r,\theta) = \frac{RC_E}{\varepsilon_0} \sum_{n=1}^{\infty} \lambda_n \sin(n\theta) \int_0^{\infty} J_n(kr) J_n(kR) \frac{\mathrm{d}k}{k} \qquad [A16]$$

where we have defined the projected multipole charge densities as

$$\lambda_n = \int_{-d}^{d} \sigma_n(z_S) \mathrm{d}z_S \qquad [A17]$$

The integral along *k* yields

$$\int_0^{\infty} J_n(kr) J_n(kR) \frac{\mathrm{d}k}{k} = \frac{1}{2n} \left(\frac{r_<}{r_>}\right)^n \qquad [A18]$$

where $r_{<>}$ is the smallest (largest) between *r* and *R*, respectively. With this, we arrive at the phase multipole expansion

$$\varphi(r,\theta) = \sum_{n=1}^{\infty} \varphi_n \left(\frac{r_<}{r_>}\right)^n \sin(n\theta) \qquad [A19]$$

where

$$\varphi_n = \frac{RC_E}{2n\varepsilon_0} \lambda_n \qquad [A20]$$

The phase multipole expansion [A19] could be deduced directly by its Laplacian nature. In fact, as argued previously in the main text, in absence of charges along the beam path the phase shift satisfies the 2D Laplace equation $\nabla_{xy}^2 \varphi = 0$, whose solution in the geometry considered can be generally expressed as [A19]. However, the multipole phase coefficients $\varphi_n$ would remain undetermined, while the treatment developed so far has established a clear and simple connection between $\varphi_n$ and the projected charge density coefficients $\lambda_n$. Now, if we could control the phase plate in terms of charge, we would be satisfied by equation [A20], as any sought phase distribution specified by $\varphi_n$ would find a counterpart in an appropriate set of $\lambda_n$ that generate it. Yet, any electrostatic device is operated by selecting voltages, not charges. Therefore, we need to develop the treatment further until we find a connection between $\varphi_n$ and $V_n$.

To develop such connection, we begin by considering the artificial scenario where the charge density coefficients are not a function of altitude on the cylindrical shell: we imagine injecting some charge uniformly along *z*, with an angular distribution dictated by the multipoles we need and freeze it in place. In this case, the projected charge coefficient are just $\lambda_n = t\sigma_n$, and the potential on the shell becomes

$$V(R,\theta,|z|<d) = \frac{R}{2\varepsilon_0} \sum_{n=1}^{\infty} \sigma_n \sin(n\theta) \int_0^{\infty} J_n^2(kR) \mathrm{d}k \int_{-d}^{d} e^{-k|z-z_S|} \mathrm{d}z_S$$

$$= \frac{R}{\varepsilon_0} \sum_{n=1}^{\infty} \sigma_n \sin(n\theta) \int_0^{\infty} J_n^2(kR) \left(1 - e^{-kd} \cosh kz\right) \frac{dk}{k}$$

$$= \frac{R}{\varepsilon_0} \sum_{n=1}^{\infty} \sigma_n \sin(n\theta) \left(\frac{1}{2n} - \int_0^{\infty} J_n^2(kR) e^{-kd} \cosh kz \frac{dk}{k}\right)$$

The remaining integral could be expressed through hypergeometrics, but its analytical form bears little interest. More instructive is a plot of the potential coefficients obtained in this way

$$V_n(z) = \frac{R\sigma_n}{\varepsilon_0} \left(\frac{1}{2n} - \int_0^{\infty} J_n^2(kR) e^{-kd} \cosh kz \frac{dk}{k}\right)$$

that represents the vertical variation of the potential along the shell. We note that the higher the multipole order, the flatter is the potential. This implies that the frozen charge approximation is more and more accurate as the multipole order increases. The exact solution of the problem would feature potential coefficients that are independent of *z*, with a geometrical dependence upon the cylinder aspect ratio τ=d/R alone. The ratio $\sigma_n/V_n$ can be interpreted as a generalized multipole capacitance per unit area.

To proceed further, we follow the physics of what would happen in reality if we unfroze the charge distribution and let it relax: it would spread out from the midplane of the shell accumulating at the rims, to lift the downwards trend of the potential profiles shown in Fig A2.2 until they all flatten at any multiple order, although the value of the equipotential would be different at each order. Since we cannot calculate the actual generalized capacitances, because doing so requires knowledge of the final charge densities after relaxation to equipotential equilibrium, we attempt a crude approximation of them: for each multipole order, we estimate the equipotential value that would be reached if the charge were allowed to spread, as the averaged potential along the shell that is established by the frozen uniform charge. This approximation, as evidenced in Fig. A2.2 is poor for the lower-order multipoles and small shell thicknesses, but it quickly becomes reasonable as *n* or τ increases. With the substitution *q=kR* we obtain:

$$\bar{V}_n = \frac{R\sigma_n}{\varepsilon_0} \left(\frac{1}{2n} - \frac{1}{t} \int_0^{\infty} J_n^2(kR) e^{-kd} \frac{dk}{k} \int_{-d}^{d} \cosh kz \, dz\right)$$

$$\bar{V}_n = \frac{\lambda_n}{4\tau\varepsilon_0 n} \left(1 - \frac{n}{\tau} \int_0^{\infty} J_n^2(q) \left(1 - e^{-2q\tau}\right) \frac{dq}{q^2}\right) \quad \text{[A21]}$$

that once again can be expressed in terms of hypergeometric, reducing to combinations of polynomials and elliptic integrals for integer *n* as in our case.

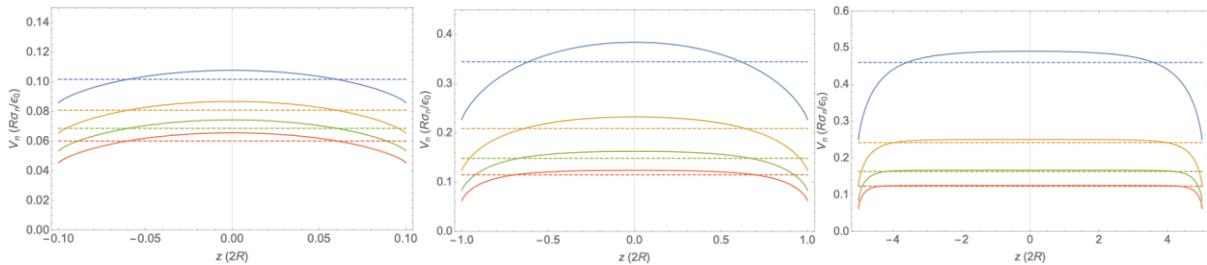

***Fig A2.2***: *Potential profile along the shell vertical coordinate generated by a frozen uniform charge (continuous lines) and their averages (dashed lines). Colors represent the multipole orders: n=1 (blue), n=2 (orange), n=3 (green), n=4 (red). The aspect ratio of the cylindrical shell varies from τ=0.1 (left), to τ=1 (middle), to τ=5 (right).*

We are now in the position to understand how the phase multipole coefficients are linked to the potential coefficients, and through them to the geometry of the device. Combining A20 and A21 we obtain

$$\varphi_n = C_E t \bar{V}_n f_n(\tau) = 2RC_E \bar{V}_n \tau f_n(\tau) = 2RC_E \bar{V}_n \tau_n(\tau) \quad \text{[A22]}$$

with

$$f_n(\tau) = \left(1 - \frac{n}{\tau} \int_0^\infty J_n^2(q) \left(1 - e^{-2q\tau}\right) \frac{dq}{q^2}\right)^{-1} \quad \text{[A23]}$$

Remarkably, the phase and potential are linked through the usual $C_E tV$, modulated by functions $f_n$ that capture the influence of both the shell geometry and the fringing fields. The multipole-dependent aspect ratio $\tau_n(\tau)=\tau f_n(\tau)$ can be interpreted as the effective thickness over which we can integrate the potential *n*-multipole to obtain the phase n-multipole. The first four phase multipole coefficients are plotted in fig A2.3.

We briefly examine the relevant limiting configuration of thick shell. In this case, the exponential goes to zero, the integral of the $J_n^2$ is finite, so that the whole integral becomes negligible and $f_n$ goes to 1: the phase is just a *t*-thick slice of the uniform (along *z*) potential. Expanding at first order in $1/\tau$ we obtain

$$f_n(\tau \to \infty) \sim 1 + \frac{4n}{(4n^2-1)\pi\tau} \quad \text{[A24]}$$

and further

$$f_{n\to\infty}(\tau \to \infty) \sim 1 + \frac{1}{n\pi\tau} = 1 + \frac{R}{nct} \quad \text{[A25]}$$

Where the constant *c* equals π/2 in this approximate scenario. This value is relatively close to unity but it is reasonable to expect that the specific value of *c* depends on the specific approximations.

The opposite limit of thin shell is significantly more challenging to obtain, since it involves logarithms and will not be treated explicitly.

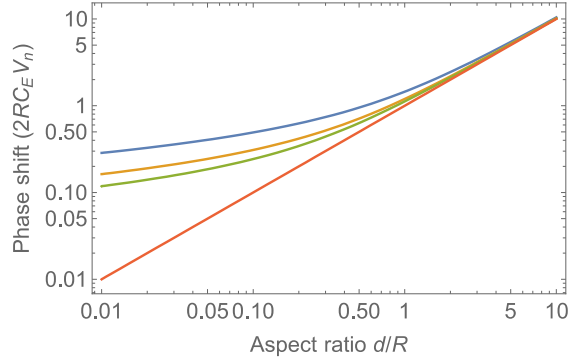

*Figure A2.3: multipole phase coefficients n=1 (blue), 2 (orange), 3 (green) and 20 (red; corresponding to the thick electrode case where the effective thickness coincides with the proper thickness) as a function of the cylindrical shell aspect ratio.*

### A3 Explicit solution for the Spiral phase plate

The spiral phase plate does not directly belong to this scheme since the device is comprised of two parts: the chopstick itself and the electrodes on the boundary conditions.

The assumption that we make is that the respective mutual induction between the boundary and the chopstick is negligible so that the two problems can be solved separately.

We will tackle here the boundary condition while using the finite element in appendix to join the two solutions. In this case of the boundary conditions of the spiral phase plate with $\varphi_{aim} = \ell\theta$

For the ideal potential $V(\theta) = \ell\theta/(C_E t)$ the Fourier coefficient can be given analytically as in polar coordinates the function $f$ is just a linear ramp. In fact, the Fourier series of a sawtooth wave has the analytical form $A_n = \ell/n$

That means the actual potentials to be applied have the form:

$$V(\theta) = \frac{1}{C_E t} \sum_m \frac{\ell/n}{1+\frac{R}{nct}} R^n \sin(n\theta + p_n) \quad [A26]$$

The result is the green curve showing a strong gradient in proximity of the chopsticks electrodes ($\theta \approx \pm\pi$) and a weaker linear ramp in front of the electrodes.

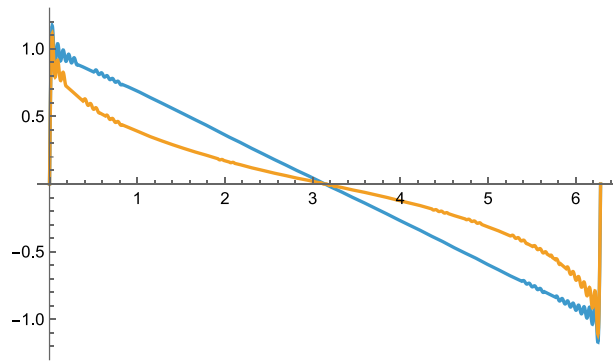

*Fig A3.1 (blue) Azimuthal distribution voltage on the boundary condition for an ideal (thick) spiral phase plate, (yellow) actual potential profile obtained with a thin electrode*

### A4 A better parametrization of the electrodes

For the testing of the device we wrote specialized software that simplifies the 8 bias to 4 main parameters that are here summarized.

**M**=main bias on the single chopstick

**O**=offset additional bias to compensate chopstick asymmetry

**F**=factor defining the relation from the chopstick to boundary conditions

**S**=overshoot additional corrections for the thin electrodes case

**B=M x F** scale factor of the boundary condition

We add a more complete parametrization of the controls of the electrodes that include also the tilt and unidirectional twofold astigmatism.

These commands can be usually substituted by the equivalent command of the microscopes.

Occasionally it can be useful to act directly on these parameters as this can be more efficient in terms of automation and control: a single controller could self-consistently act on all parameters of the vortex. The 3 additional parameters are

**A**=astigmatism (single axis along the diagonal xy)

**Tx, Ty**=tilt along x and y

$$\varphi = A\cos(2\theta + \tfrac{\pi}{4}), \varphi = T_x \cos(\theta), \varphi = T_y \sin(\theta)$$

| electrode | Formula vs parameters | |
|---|---|---|
| V1 | -M +O | + A 0.7 + 1.4Ty |
| V2 | -0.5 B | + A 0.7 + 1.4 Ty |
| V3 | (-0.5+1/11)x(S+B) | +A +Tx+Ty |
| V4 | (-0.5+4/11)x(S+B) | -A +Tx-Ty |
| V5 | (-0.5+7/11)x(S+B) | +A -Tx-Ty |
| V6 | (-0.5+10/11)x(S+B) | -A -Tx-Ty |
| V7 | 0.5 B | + A 0.7 + 1.4 Ty |
| V8 | M +O | + A 0.7 + 1.4 Ty |

### A5 Simulation with a realistic substrate

In this appendix we report the calculation for a more realistic simulation with grounded substrate. The grounded substrate is a structure that is directly attached to the end of the main MEMS/multipole structure: it produces as an effect the reduction of the fringing field that is therefore present on only one side. This can be roughly accounted for by changing scale factor c of the correction.

Figure A5.1 shows the comparison between the simplified simulation (no substrate) and the more realistic case. As visible, the correction is significantly reduced while remaining proportional to the case of freestanding ring without substrate.

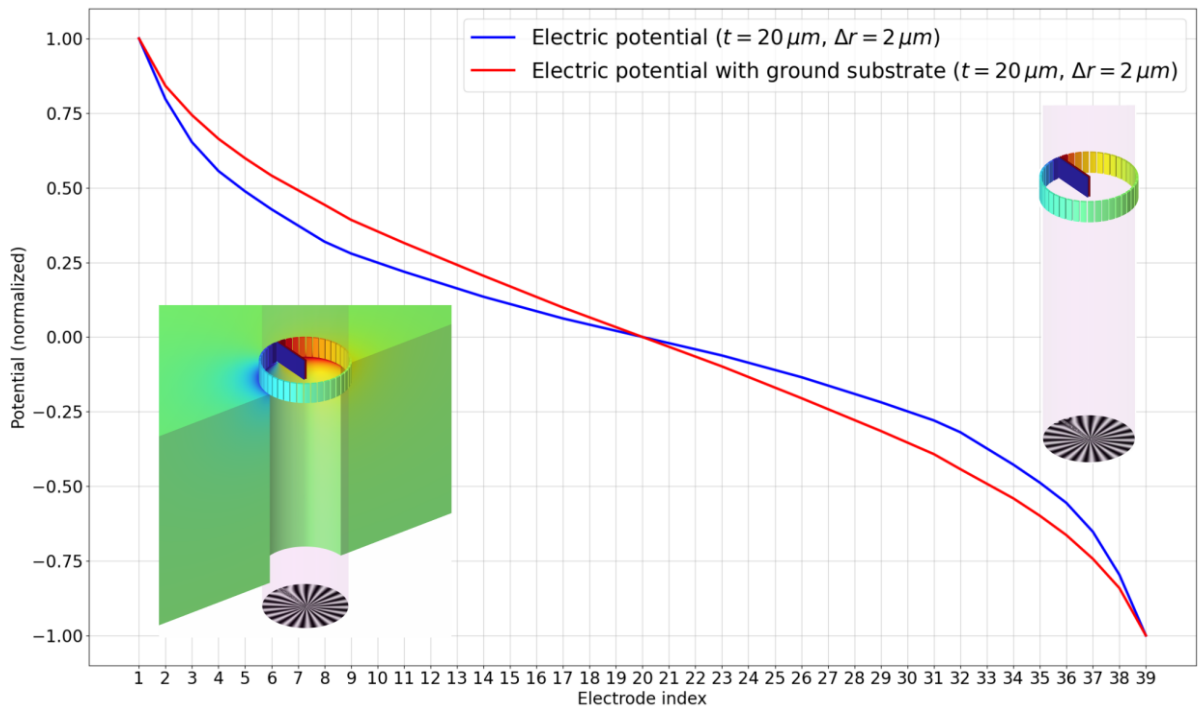

**Fig A5.1** Comparison of the applied potential to get a perfect vortex beam considering an isolated thin ring of electrodes and one physically connected to a grounded substrate.